\documentclass[preprint,5p,times,twocolumn,12pt]{elsarticle-mod}

\usepackage{ifpdf}
\usepackage{graphicx,natbib,lineno,color}
\ifpdf
\usepackage{amssymb}
\usepackage{lipsum}
\usepackage{bm}
\usepackage{amssymb}
\usepackage{pifont}
\usepackage{slashed}

\usepackage[%
breaklinks=true,%
colorlinks=true,%
linkcolor=blue,anchorcolor=blue,%
citecolor=blue,filecolor=blue,%
menucolor=blue,pagecolor=blue,%
urlcolor=blue]{hyperref}

\makeatletter
\def\elsartstyle{%
	\def\normalsize{\@setfontsize\normalsize\@xiipt{14.5}}
	\def\small{\@setfontsize\small\@xipt{13.6}}
	\let\footnotesize=\small
	\def\large{\@setfontsize\large\@xivpt{18}}
	\def\Large{\@setfontsize\Large\@xviipt{22}}
	\skip\@mpfootins = 18\p@ \@plus 2\p@
	\normalsize
}
\@ifundefined{square}{}{\let\Box\square}
\makeatother

\newcommand{\nn}{\nonumber}
\def\d{{\mathrm{d}}}

\def\tr{{\mathrm{tr}}}
\def\Hilbert{{\mathcal{H}}}
\def\H{{\mathrm{H}}}
\def\R{{\mathrm{R}}}
\def\E{{\mathrm{E}}}

\def\tr{{\mathrm{tr}}}

\journal{Physics Letters {\bf B776} (2018) 10--16}
\def\date{{8 November 2017; LaTeX-ed \today}}

\begin{document}
	
\begin{frontmatter}
	
\title{Entropy/information flux in Hawking radiation}

\author{Ana Alonso--Serrano}

\address{Institute of Theoretical Physics, Faculty of Mathematics and Physics,
		Charles University, 18000 Prague, Czech Republic }

\ead{a.alonso.serrano@utf.mff.cuni.cz\break}  

\author{Matt Visser}

\address{School of Mathematics and Statistics,
Victoria University of Wellington;
PO Box 600, Wellington 6140, New Zealand.}

\ead{matt.visser@sms.vuw.ac.nz\break} 

\begin{abstract}
Blackbody radiation contains (on average) an entropy of $3.9\pm2.5$ bits per photon. If the emission process is unitary, then this entropy is exactly compensated by ``hidden information'' in the correlations.  We extend this argument to the Hawking radiation from GR black holes, demonstrating that  the assumption of unitarity leads to a perfectly reasonable entropy/information budget. The key technical aspect of our calculation is  a variant of the ``average subsystem'' approach developed by Page, which we extend beyond bipartite pure systems, to a tripartite pure system that considers the influence of the environment.

\medskip\noindent
\emph{Preprinted as:}  arXiv: 1512.01890 [gr-qc]

\medskip\noindent
\emph{Published as:} Physics Letters {\bf B776} (2018) 10--16; doi: https://doi.org/10.1016/j.physletb.2017.11.020
 \end{abstract}

\begin{keyword}
	Black holes. Entropy. Information. Average subsystem	
\end{keyword}

\end{frontmatter}

\section{Introduction}\label{S:intro}

The ``information puzzle'' due to  the Hawking evaporation of GR black holes continues to provoke much heated discussion and debate~\cite{info-puzzle,firewall,apologia,Chen:2015,Mathur:2015,Nomura:2012a,Nomura:2012b,Israel:2014,Albrecht:2014,w/o-loss,Mathur:2009,Preskill:1992,Ashtekar:2005,Hayward:2005,Hayward:2005b,Hayward:2005c,Bardeen:2014,Hawking:2015}.
On the other hand, there simply is no ``information puzzle'' associated with chemical burning~\cite{burning}, nor with the Hawking radiation from \emph{analogue} black holes~\cite{Unruh:1981,Visser:1993,Visser:1998,Visser:1998b,Visser:2001a,LRR,Barcelo:2006,Visser:2007,Visser:2010,Lake-Como,Weinfurtner:2010,Weinfurtner:2013,Unruh:amplifier,Unruh:measured,Steinhauer:2014,Steinhauer:2015a,Steinhauer:2015b,Belgiorno:2010,Schutzhold:2010,Belgiorno:2010b,Rubino:2011,Liberati:2011}, where  physics is manifestly unitary. 
(Horizons, if present at all, are \emph{apparent/trapping} horizons;  definitely not \emph{event} horizons~\cite{Visser:1997,Barcelo:2004,trapped}.) 
Previously we carefully analyzed the blackbody radiation from a ``blackbody furnace''~\cite{burning}. In the current article we focus on Hawking radiation from both \emph{analogue} and GR black holes. Despite many claims to the contrary,  (assuming unitarity and complete evaporation), the Hawking evaporation process is relatively benign, no worse than burning a lump of coal.

\section{Entropy/information in blackbody radiation}

When burning a lump of coal (or an encyclopaedia for that matter) in a blackbody furnace, individual photons in the resulting blackbody radiation carry (on average) an entropy/ information content of~\cite{burning}
\begin{equation}
\langle\hat S_2\rangle \approx \; 3.90 \pm 2.52 \;  \hbox{ bits/photon}.
\end{equation}
We use $S$ to denote the physical entropy, $\hat S = S/k_B$ for the dimensionless entropy measured in nats (natural units), and $\hat S_2 =  \hat S/\ln2$. We now apply these results within the context of Hawking radiation, paying particular attention to the von~Neumann entanglement entropy, and thence to the Page curve,  as one of the main features underlying the firewall argument. 

\section{Hawking evaporation: \\
analogue and GR black holes}
 
\emph{Analogue} black holes \cite{Unruh:1981,Visser:1993,Visser:1998,Visser:1998b,Visser:2001a,LRR,Barcelo:2006,Visser:2007,Visser:2010,Lake-Como} have unitary Hawking flux; the relevant \emph{blocking/acoustic} horizons are \emph{apparent/trapping} horizons. Hawking quanta simply deliver a coarse-grained thermodynamic entropy~\cite{ana:entropy} $S = \hbar\,\omega/T$ to the radiation field; exactly compensated by the information hidden in the correlations between the quanta~\cite{burning}. \emph{Analogue} black holes provide the  only \emph{experimental} evidence for the reality of Hawking radiation~\cite{Weinfurtner:2010,Weinfurtner:2013,Unruh:amplifier,Unruh:measured,Steinhauer:2014,Steinhauer:2015a,Steinhauer:2015b}, showing a quite standard unitary preserving quantum physics without involving any ``information puzzle''.

If there is any ``information puzzle'' for GR black holes, it is not Hawking radiation \emph{per se} that is the central issue. It is the assumed existence of \emph{event} horizons (which certainly do exist in the classical limit) surviving in the semiclassical quantum realm that is the source of the potential difficulties.

The ``information puzzle'' can be traced back to Hawking's 1976 article~\cite{info-puzzle} where he introduced the concept of ``hidden surface'', which was  to be understood as a synonym for ``absolute causal horizon''. 
(Hawking has since twice abjured the semiclassical survival of event horizons~\cite{dublin,weather}.) Also note that \emph{event} horizons are simply not physically observable (in any finite size laboratory), whereas \emph{apparent/trapping} horizons certainly are physically observable, at least in spherical symmetry~\cite{observability}. Furthermore, even in a general relativity context, \emph{event} horizons are simply not essential for generating a Hawking-like flux~\cite{Hajicek:1986hn,Visser:2001,Barcelo:2010xx,Barcelo:2010yy}.

\vspace{-10pt}
\section{Thermodynamic entropy: \\Hawking flux from a GR black hole}

For the Hawking evaporation of a GR black hole, we shall argue that classical thermodynamic entropy fluxes stay the same. Quantum entanglement entropy fluxes \emph{might} in principle differ; that is essentially what all the arguing is about. To clarify these issues we shall compare and contrast the behaviour of the classical thermodynamic (Clausius) entropy with the quantum entanglement (von~Neumann) entropy (of suitably defined subsystems).

\subsection{Loss of Bekenstein entropy of the GR black hole}

Let us first estimate the Bekenstein entropy loss of the black hole per emitted quanta. We assume for simplicity an exact Planck spectrum at the Hawking temperature, this being a good zeroth-order approximation to the actual physics~\cite{sparsity,thermality}. 
For a Schwarzschild black hole we have
\begin{eqnarray}
{\d S\over \d N} &=&  {\d S/\d t\over \d N/\d t} =
(8\pi k_B G M/\hbar c) (\hbar \langle\omega\rangle/c^2 ),
\end{eqnarray}
where so far we have only used the definition of Bekenstein entropy and the conservation of energy.
Thus, for a Planck spectrum of emitted particles~\cite{burning}
\begin{equation}
{\d S\over \d N} ={k_B  \pi^4\over30\;\zeta(3)}. 
\end{equation}
This is Bekenstein entropy loss of the black hole (per emitted massless boson).

\subsection{Gain of thermodynamic entropy of the radiation}

\begin{figure}[!h]
	\begin{center}
		\includegraphics[scale=0.55]{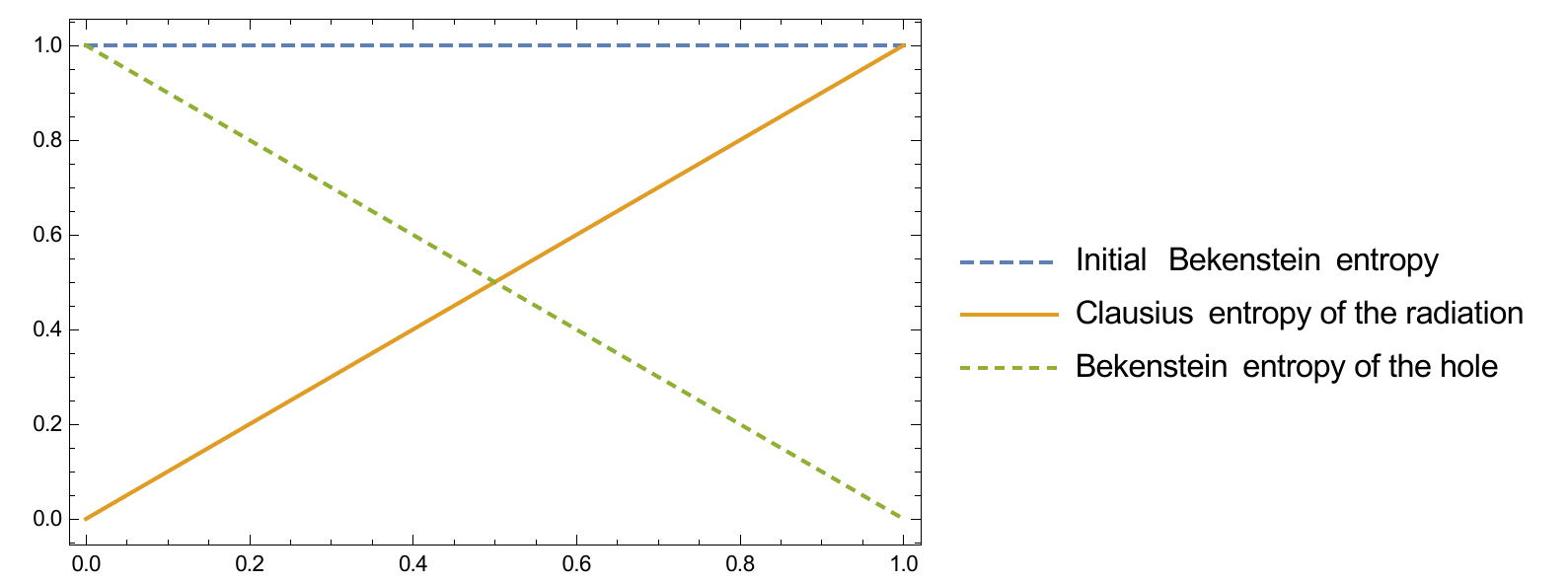}
		\caption{{\bf Clausius (thermodynamic) entropy balance:} \newline
			As the black hole Bekenstein entropy (defined in terms of the area of the horizon) decreases the Clausius entropy of the radiation increases to keep total entropy constant and equal to the initial Bekenstein entropy.}
		\label{F:entropy-balance1}
	\end{center}
\end{figure}

Contrast this with the thermodynamic entropy gain (Clausius entropy gain) of the external radiation field (the Hawking flux) per emitted quanta. We have 
\begin{equation}
{\d S\over \d N} = {\d E/T_H\over \d N}  = {\hbar\langle\omega\rangle \over T_H} 
 =  {k_B \pi^4\over30\;\zeta(3)}.
\end{equation}
Independent of the details of the microphysics, at the macroscopic level the Hawking radiation is essentially just (adiabatically) transferring the Bekenstein entropy from the black hole into the Clausius entropy of the radiation field; there are no significant qualifications or limitations to this result. Throughout the evaporation process, in terms of the initial Bekenstein entropy $ S_\mathrm{Bekenstein,0}$ we have (see Fig.~\ref{F:entropy-balance1}):
\begin{equation}
\label{E:clausius}
S_\mathrm{Bekenstein}(t) + S_\mathrm{Clausius}(t) = S_\mathrm{Bekenstein,0}.
\end{equation}

\subsection{Total number of emitted Hawking quanta in GR}

As a cross-check, let us estimate the total number of emitted massless quanta. We have
\begin{equation}
{\d N\over \d M} =   {(\d N/\d t)\over (\hbar \langle\omega\rangle/c^2) (\d N/\d t)} 
={30 \zeta(3)\over\pi^4} {8\pi GM\over\hbar c}.
\end{equation}
Integrating this we have:
\begin{equation}
N = 
{30 \zeta(3)\over\pi^4}\; \hat S 
\;\approx\; 0.26  \; \hat S_2.
\end{equation}
The total number of emitted quanta is proportional to the original Bekenstein entropy. Conversely:
\begin{equation}
{\d\hat S_2\over \d N}=  {\pi^4\over30 \zeta(3)\ln 2 } \approx \;3.90 \;  \hbox{ bits}.
\end{equation}
Semi-classically (at the level of macroscopic thermodynamics) everything holds together very well; the total number of massless quanta emitted over the life of the black hole is comparable to the (initial) dimensionless Bekenstein entropy.

\vspace{-10pt}
\section{Entanglement entropy: \\
Hawking flux from a GR black hole}

Now we come to the heart of the matter: Do these classical thermodynamic entropy arguments match with quantum entropy arguments based on the von~Neumann entropy? If we wish to preserve unitarity, then over the lifetime of the black hole we will have to encode approximately $3.9\pm2.5$ bits per photon of hidden information into the Hawking flux.
But, can we implement this ``purification''  process ``continuously'', or is it all hidden in a (non-perturbative) burst of information at/near total evaporation? 
Or after the so-called Page time?~\cite{Page-curve}. We argue, assuming unitarity, complete evaporation, and a variant of the ``average subsystem'' argument, that the purification process is continuous and ongoing.

\subsection{Entanglement: Subsystem entropies}

Page~\cite{Page:subsystem} has established a number of interesting results regarding average subsystem entropies. Consider a Hilbert space that factorizes, $\mathcal{H}_{AB} = \mathcal{H}_{A} \otimes \mathcal{H}_{B}$, and on that Hilbert space consider a pure state $\rho_{AB} = |\psi\rangle\langle\psi |$. Now define subsystem density matrices via the partial traces:  $\rho_i = \tr_j( |\psi\rangle\langle\psi |)$, where $i$, $j$ runs over $A,B$. Then the subsystem von~Neumann entanglement entropies, $\hat S_i= -\tr(\rho_i\ln\rho_i)$, both satisfy
\begin{equation}
\hat S_A = \hat S_B \leq \ln \min\{\dim(\mathcal{H}_{A}), \dim(\mathcal{H}_{B})\}.
\end{equation}
This particular equality and inequality hold \emph{before} any averaging is enforced. Page then considered the effect of taking a uniform average over all pure states on $\Hilbert_{AB}$. Taking $n_1=\dim(\Hilbert_A)$ and $n_2=\dim(\Hilbert_B)$, with $m=\min\{n_1,n_2\}$ and $M=\max\{n_1,n_2\}$,  he defined the equivalent of
\begin{equation}
\hat S_{n_1,n_2} = \langle \hat S_A \rangle = \langle \hat S_B \rangle  \leq \ln m.
\end{equation}
The central result of Ref.~\cite{Page:subsystem} is that the average subsystem entropy is extremely close to its maximum possible value. (So that the ``average subsystem'' is effectively very close to being ``maximally mixed''.) Combined with the exact result derived by Sen~\cite{Sen}, (Sen provided a formal analytic proof of an exact result conjectured by  Page), and our own calculations involving an expansion in terms of the harmonic numbers~\cite{ana:pra},  this can be strengthened to a strict bound
\begin{equation}
\hat S_{n_1,n_2} = \langle \hat S_A \rangle = \langle \hat S_B \rangle  
\in \left( \ln m - {\textstyle{1\over2}}, \ln m\right).
\end{equation}
The average subsystem entropy is within ${1\over2}$ nat, (less than ${1\over2\ln2} <{3\over4}$ bit), of its maximum possible value.

\subsection{Bipartite entanglement: \\ GR black hole + Hawking radiation}

In Ref.~\cite{Page-curve} Page applies the average subsystem formalism to an idealized bipartite system consisting solely of (GR black hole)+(Hawking radiation). This is a ``closed box'' argument, ignoring the rest of the universe. In the idealized bipartite HR system, initially there is not yet any Hawking radiation,  $\Hilbert_\mathrm{Hawking\,radiation} = \Hilbert_\R$ is trivial, (so it is 1-dimensional), while  $\Hilbert_\mathrm{black\,hole}= \Hilbert_\H$ is enormous. But it is the minimum dimensionality that dominates the average subsystem entropy  and so $(\hat S_{n_\H,n_\R})_0=0$.
We shall use a subscript 0 to denote time zero.  Likewise a subscript $\infty$ will denote time infinity. After the black hole has (by assumption) completely evaporated it is  $\Hilbert_\mathrm{black\,hole}= \Hilbert_\H$ that is trivial (1-dimensional), and so $(\hat S_{n_\H,n_\R})_\infty=0$. At intermediate times both $\Hilbert_\H$ and $\Hilbert_\R$ are nontrivial, (having dimensionality greater than unity), so the average subsystem entropy is non-zero. 

\begin{figure}[!h]
	\begin{center}
		\includegraphics[scale=0.55]{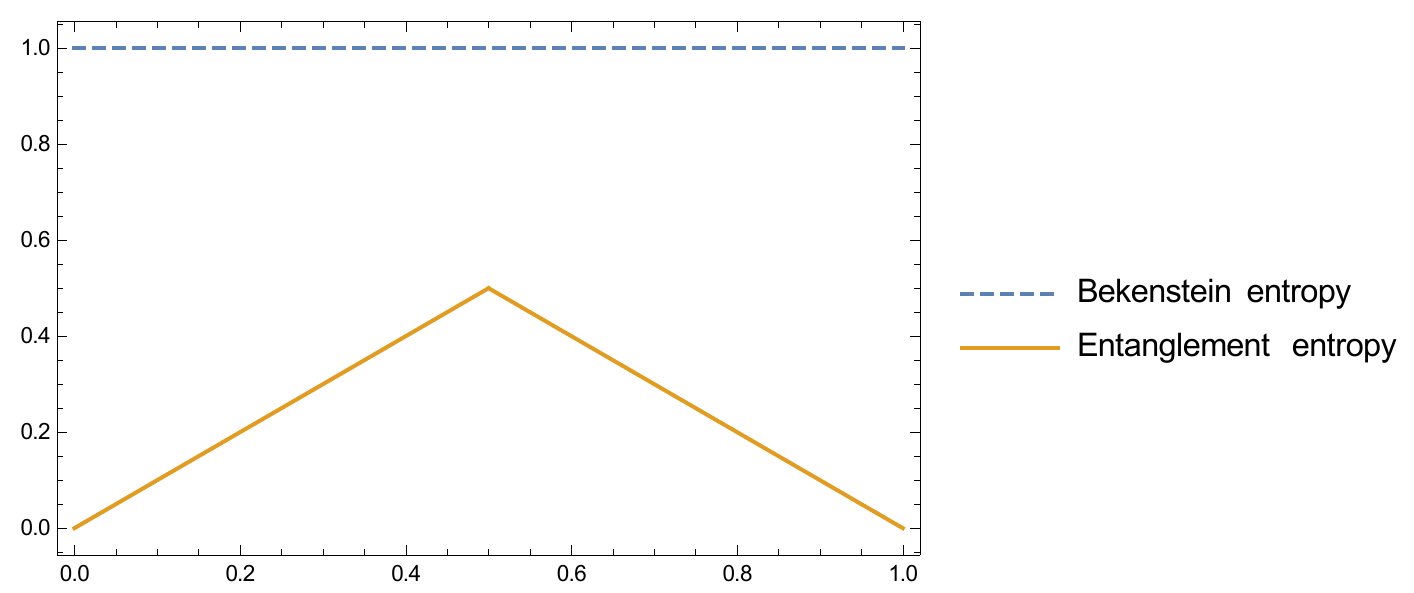}
		\caption{{\bf Page curve, bipartite entanglement entropy:} \newline Under the ``average subsystem'' assumption applied to a pure-state bipartite system consisting of (black hole) plus (Hawking radiation) the entanglement entropy rises from zero to one half the initial Bekenstein entropy before dropping back to zero. }
				\label{F:Page-curve}
	\end{center}
\end{figure}

Since the evolution is assumed unitary the dimensionality of the total Hilbert space is constant, so 
\begin{equation}
n_\H(t) \; n_\R(t) = n_\mathrm{H_0} = n_\mathrm{R_\infty}. 
\end{equation}
The subsystem entropy rises from zero to some maximum and then descends back to zero. 
That maximum is reached when
\begin{equation}
\hat S_{n_\H,n_\R}(t=t_\mathrm{Page})\approx {1\over2} \; \ln  n_\mathrm{H_0}. 
\end{equation}
It is the (symmetric) sawtooth shape of the Page curve (see Fig.~\ref{F:Page-curve}) that underlies much of the modern discussion surrounding the ``information puzzle'', and in particular the asserted and much debated existence of firewalls~\cite{firewall,apologia,Chen:2015,Mathur:2015,Nomura:2012a,Nomura:2012b,Israel:2014,Albrecht:2014}.  But is there some way of evading the current argument? 

One particularly disturbing feature of the current  bipartite argument is that the subsystem entropy is initially zero. But this observation is in marked tension with the fact that the Bekenstein entropy of the black hole is initially enormous, and this Bekenstein entropy is usually attributed to some form of entanglement entropy. In this model the Bekenstein entropy is \emph{never} the entanglement entropy of the black hole, it is instead the maximum entropy that the black hole \emph{could have had} given the size of the Hilbert space used to describe the bipartite system.

Page's main result implies that the black hole subsystem is maximally entangled with the radiation  subsystem. But, at the same time, if we sub-divide the Hawking radiation subsystem between early 
and late radiation (respectively, before and after Page time), these two subsystems would be also maximally entangled with each other, and also with the black hole subsystem. One way of looking at the problem relies on the fact that due to the monogamy of entanglement, this is not possible. This was one of the motivations for the firewall proposal~\cite{firewall,apologia}. We are much less sanguine regarding the physical relevance of the Page curve, and in fact will argue against the physical relevance of the Page curve. 

We shall instead argue that it is more appropriate to consider a tripartite system taking into account also the environment (rest of the universe),
and that the average subsystem entropy argument, when  applied to this (pure state) tripartite system, yields much more acceptable (and hopefully noncontroversial) physics. 

\subsection{Bipartite entanglement: \\
Asymmetric subsystem information}

\begin{figure}[!h]
	\begin{center}
		\includegraphics[scale=0.55]{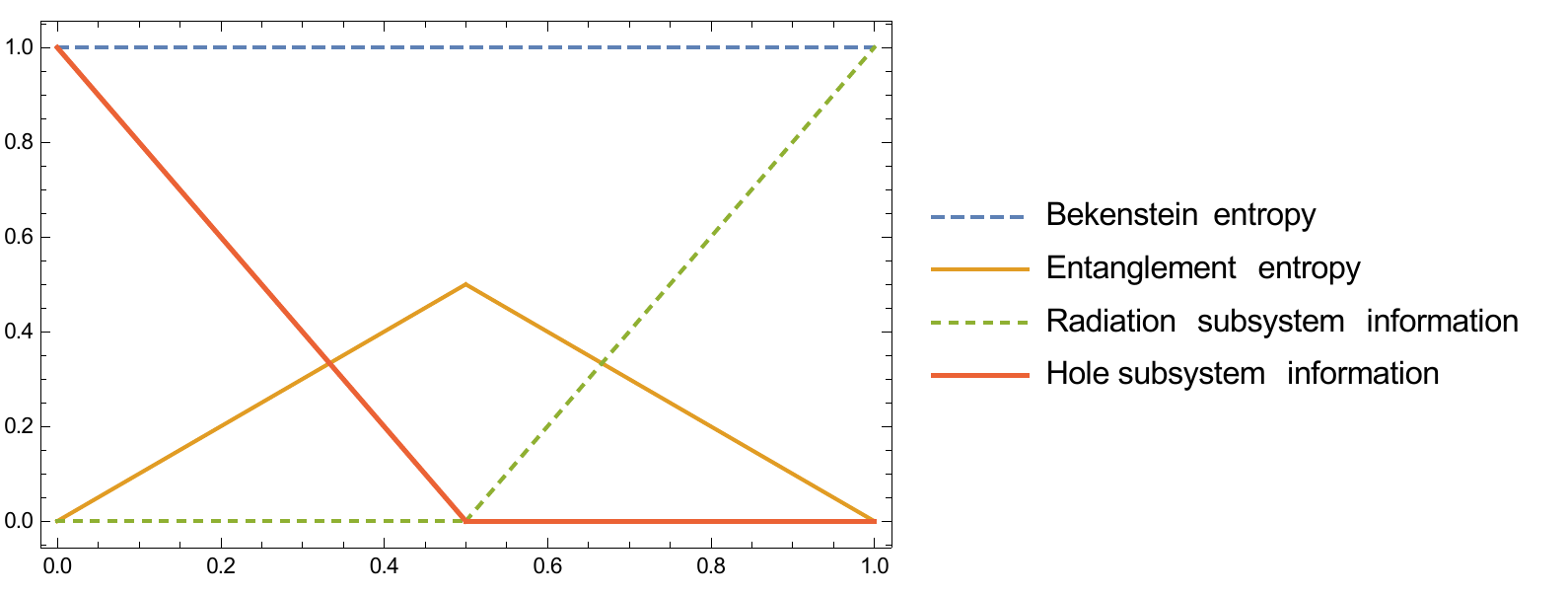}
		\caption{{\bf Page curves for entanglement entropy and (asymmetric) subsystem information:} Note the ``kinked'' behaviour of the (asymmetric) subsystem information and that the ``not-quite sum rule'' $\langle\tilde I_{\H,\R}\rangle + \langle\tilde I_{\R,\H}\rangle+ 2 \langle\hat S_\H \rangle = \hat S_\mathrm{Bekenstein,0}$ is satisfied.}
		\label{F:Page-summary-standard}
	\end{center}
\end{figure}

In Ref.~\cite{Page-curve} Page also defines a novel asymmetric version of subsystem information, (whereas the subsystem information of Ref.~\cite{Page:subsystem} is symmetric):
\begin{equation}
\!\!\!
\tilde I_{n_1,n_2}  = \ln n_1  - \hat S_{n_1,n_2};
\quad
\tilde I_{n_2,n_1} = \ln n_2  - \hat S_{n_1,n_2}.
\quad
\end{equation}
This definition of asymmetric subsystem information is not entirely standard, and we shall see that its physical interpretation is not entirely clear. 
Nevertheless, approximately (to within ${1\over2}$ nat)  the ``random subsystem'' argument leads to
\begin{equation}
\tilde I_{n_1,n_2} \approx\ln\left(n_1\over m\right); 
\quad
\tilde I_{n_2,n_1} \approx\ln\left(n_2\over m\right).
\end{equation}
This quickly leads to the subsystem information version of the Page curve, see Fig.~\ref{F:Page-summary-standard}.

If one only considers the bipartite system of black hole and Hawking radiation, (ignoring the rest of the universe), then (assuming the black hole is initially in some unknown pure state) the asymmetric subsystem information exhibits odd features as sketched in Fig.~\ref{F:Page-summary-standard}.  By construction this bipartite system satisfies the ``not-quite sum rule''
\begin{equation}
\langle\tilde I_{\H,\R}\rangle +  \langle\tilde I_{\R,\H}\rangle + 2 \langle\hat S_\H \rangle 
= 
\hat S_\mathrm{Bekenstein,0}. 
\end{equation}
(We say ``not-quite sum rule'' because of the annoying factor 2 in front of $\langle\hat S_\H \rangle $.)

The hole subsystem information (in this bipartite model) does not have a direct physical interpretation; is the \emph{defect} between the maximum entropy that the black hole \emph{could have had} (given the time dependent size of the black hole Hilbert space) and the entanglement entropy.

\subsection{Bipartite entanglement: Mutual information}

\begin{figure}[!h]
	\begin{center}
		\includegraphics[scale=0.55]{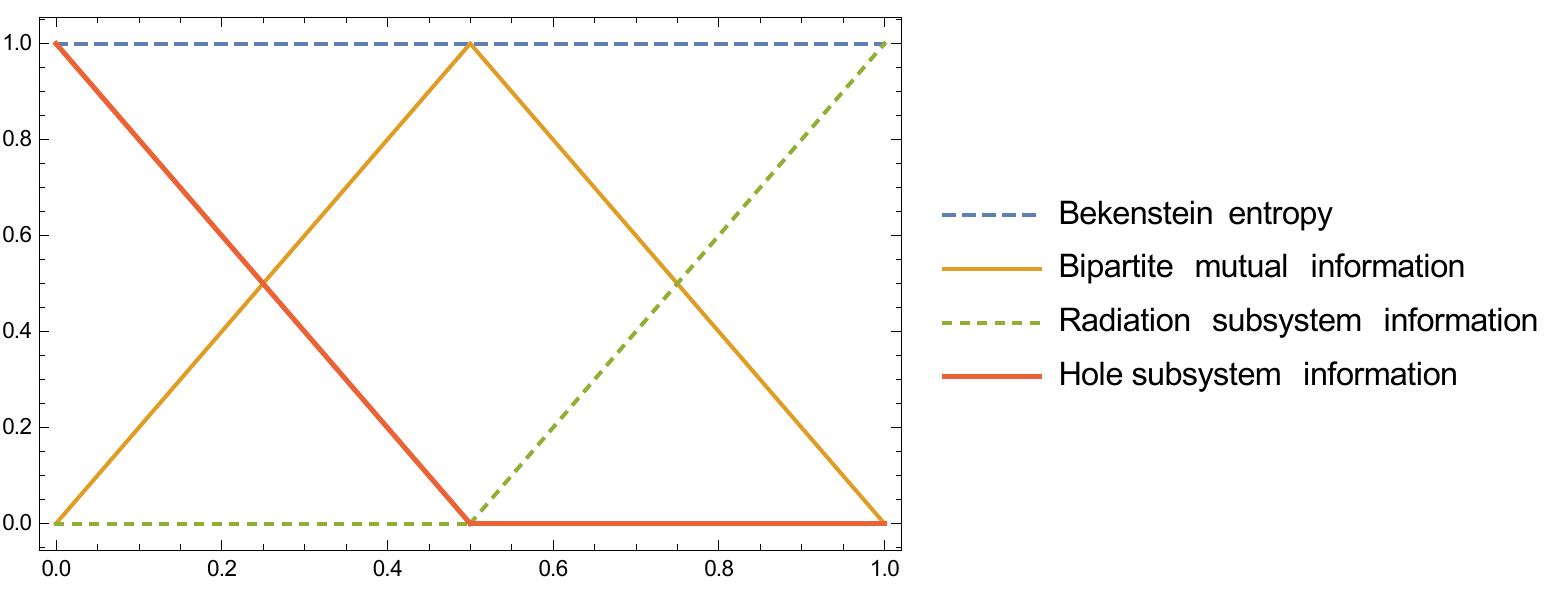}
		\caption{{\bf Modified Page curves, bipartite mutual information and  (asymmetric) subsystem information:} Note that the ``sum rule'' $\langle\tilde I_{\H,\R}\rangle + \langle\tilde I_{\R,\H}\rangle + \langle\hat I_{\H:\R}\rangle = \hat S_\mathrm{Bekenstein,0}$ is satisfied. }
		\label{F:Page-summary}
	\end{center}
\end{figure}

It should be emphasized that mutual information is certainly not the same as what Page calls the subsystem information~\cite{Page:subsystem,Page-curve}. In general one has
\begin{equation}
I_{A:B} = S_A +S_B - S_{AB}.
\end{equation}
For the bipartite HR system considered by Page one finds the particularly simple result
\begin{equation}
I_{\H:\R} = 2 S_\H = 2 S_\R.
\end{equation}
More specifically, in dimensionless units, and after applying the ``average subsystem'' argument
\begin{equation}
\langle \hat I_{\H:\R} \rangle= 2 \langle\hat S_\H \rangle = 2 \langle \hat S_\R \rangle 
\approx  2\ln\min\{n_\H,n_\R\}.
\end{equation}
While at first glance this seems uninteresting, when combined with Page's asymmetric subsystem information this leads to
the approximate sum rule
\begin{eqnarray}
&& \langle \tilde I_{\H,\R} \rangle  + \langle \tilde I_{\R,\H} \rangle + \langle \hat I_{\H:\R}  \rangle \approx 
\ln\left(n_\H n_\R\right) 
\nn \\
&& 
\qquad
\approx \ln n_{\H_0} \approx \hat S_{\mathrm{Bekenstein,0}}.
\end{eqnarray}
Here the approximation is now valid to within ${3\over2}$ nat. 
This sum rule is summarized in Fig.~\ref{F:Page-summary}.

\subsection{Tripartite entanglement:	
	GR black hole + \\Hawking radiation + rest of universe}

Consider the tripartite system with Hilbert space  $\Hilbert_\mathrm{HRE}  = \Hilbert_\H \otimes \Hilbert_\R \otimes \Hilbert_\E$; the subscripts denoting the black hole, Hawking radiation, and environment (rest of the universe). Take the entire universe to be in a pure state, so at all times $S_\mathrm{HRE}(t)=0$, while the subsystem entropies satisfy: $S_\H(t) = S_\mathrm{RE}(t)$, and $S_\R(t) = S_\mathrm{HE}(t)$, and $S_\E(t) = S_\mathrm{HR}(t)$. At time zero, as in the biparite system, $\Hilbert_{\H_0}$  is trivial (1-dimensional).  Then
\begin{equation}
S_\mathrm{H_0} = S_\mathrm{E_0};  \qquad S_\mathrm{R_0} = 0 = S_\mathrm{HE_0}.
\end{equation}
Once the black hole has completely evaporated, then  it is $\Hilbert_{\H_\infty}$ that becomes trivial. Then
\begin{equation}
S_\mathrm{H_\infty}=0= S_\mathrm{RE_\infty};  \qquad S_\mathrm{R_\infty} = S_\mathrm{E_\infty}.
\end{equation}
As the evolution is assumed unitary the total dimensionality of the Hilbert space must be fixed, but now the role of the environment is simply to give the HR subsystem something to be entangled with --- the environment does not itself directly participate in the Hawking evaporation process --- so the unitary time evolution operator factorizes as $U_\mathrm{HRE}(t) = U_\mathrm{HR}(t) \otimes U_\E(t)$. Therefore $n_\mathrm{E_0}=n_\E(t)=n_\mathrm{E_\infty} \equiv n_\E$; and $n_\H(t) \;n_\R(t) =   n_\mathrm{H_0} = n_\mathrm{R_\infty}$. That is, during the evaporation process the dimensionality of the black hole Hilbert space is being transferred to the Hawking radiation Hilbert space.

To quantify things we now make an additional assumption: That the Bekenstein entropy can be identified with the average entanglement entropy. In dimensionless units at time zero we have
\begin{equation}
\hat S_\mathrm{Bekenstein,0} = \langle \hat S_\mathrm{H_0}\rangle \approx 
\ln\min\{n_\mathrm{H_0}, n_\E \},
\end{equation}
to within ${1\over2}$ nat.  But the Bekenstein entropy depends only on intrinsic properties of the black hole, not on its environment, so we must have $\min\{n_\mathrm{H_0}, n_\E \} = n_\mathrm{H_0}$
whence $n_\mathrm{H_0} \leq n_\E.$
Subsequently, at later times we would still assert
\begin{eqnarray}
\hat S_\mathrm{Bekenstein}(t) &=& \langle \hat S_\H(t)\rangle 
\nn\\ & \approx &
\ln\min\{n_\H(t), n_\R(t) \, n_\E \}.
\end{eqnarray}
(With the approximation holding to  within ${1\over2}$ nat.)  But now note $n_\H(t) \leq n_\mathrm{H_0} \leq n_\E \leq n_\R(t) \,n_\E$. Therefore (as one would expect) 
\begin{equation}
\hat S_\mathrm{Bekenstein}(t) = \langle \hat S_\H(t)\rangle \approx  \ln n_\H(t),
\end{equation}
throughout the entire evolution. 

Conversely, for the average entanglement entropy of the radiation, (with the HE subsystem), we have
\begin{equation}
\langle \hat S_\R(t)\rangle \approx 
\ln\min\{n_\R(t), n_\H(t) \, n_\E \},
\end{equation}
But $n_\R(t) \leq n_\R(t) n_\H(t) = n_\mathrm{H_0} \leq n_\E 
\leq n_\H(t) \, n_\E$. So
\begin{equation}
\langle \hat S_\R(t)\rangle \approx 
\ln n_\R(t),
\end{equation}
throughout the entire evolution. 

\begin{figure}[!h]
	\begin{center}
		\includegraphics[scale=0.55]{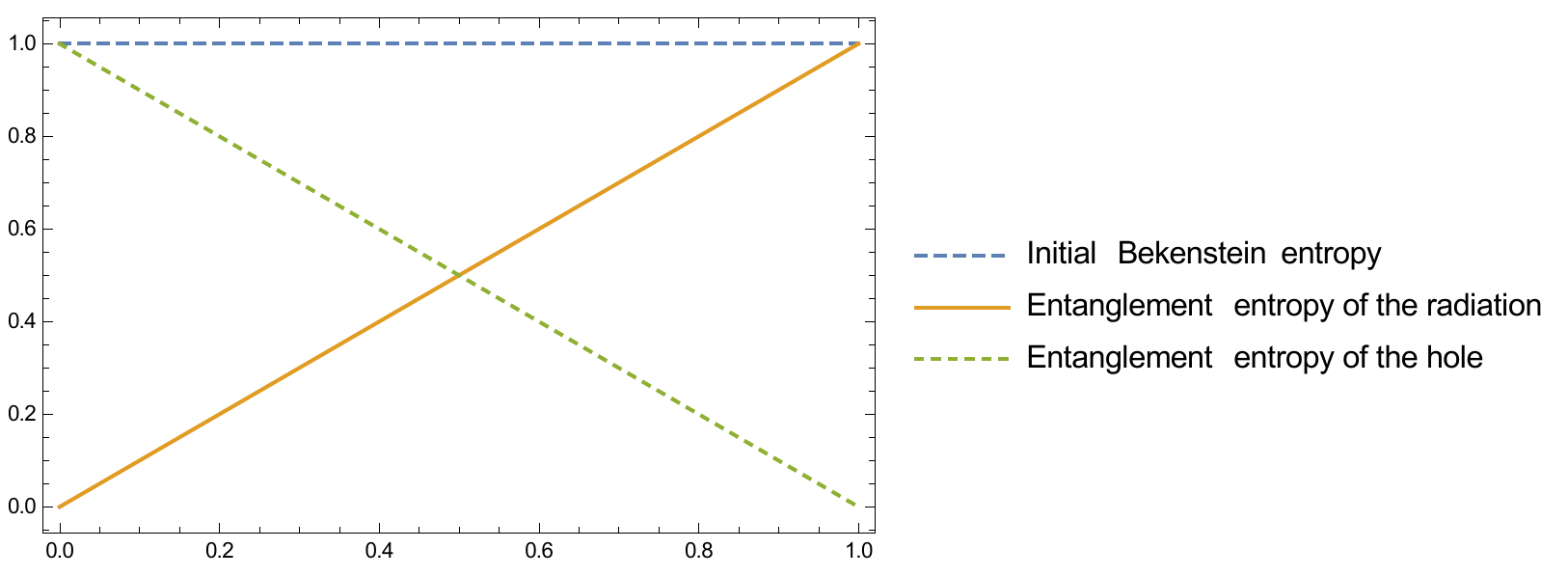}
		\caption{{\bf Tripartite quantum (von Neumann) entropy flux:}  
The quantum (von~Neumann) analysis now reproduces the Clausius (thermodynamic) analysis. As black hole Bekenstein entropy (entanglement entropy) decreases, the entanglement entropy of the radiation increases, to keep total entropy approximately constant, at least to within 1 nat. In the limit where the environment (rest of universe) becomes arbitrarily large the correspondence is exact.}
		\label{F:entropy-balance}
	\end{center}
\end{figure}

Combining these  two results
\begin{eqnarray}
&& \langle \hat S_\H(t)\rangle +\langle \hat S_\R(t)\rangle \approx 
\ln n_\H(t) + \ln n_\R(t) 
\nn \\
&&\qquad
= \ln [n_\H(t) \,n_\R(t)]
= \ln n_\mathrm{H_0}.
\end{eqnarray}
Here the approximation is valid to within at worst 1 nat. In view of earlier assumptions, we can rephrase this (to within 1 nat) as
\begin{equation}
\hspace{-10pt}
\hat S_\mathrm{Bekenstein}(t) +\langle \hat S_\mathrm{Hawking\,radiation}(t)\rangle \approx  \hat S_\mathrm{Bekenstein,0}.\;\;
\end{equation}
This now is the quantum von~Neumann entropy version of the result we previously obtained by using classical Clausius entropy arguments. (Compare with Eq.~(\ref{E:clausius}).) Note the only significant change is that the equality now holds only to within 1 nat. (See Fig.~\ref{F:entropy-balance}.)

\subsection{Tripartite entanglement:\\ 
The ``rest of the universe'' environment}

What can we say concerning the entanglement of the (black hole)+(Hawking radiation) subsystem with the rest of the universe? We have
\begin{equation}
\hspace{-10pt}
\langle \hat S_\E(t)\rangle=  \langle \hat S_\mathrm{HR}(t)\rangle\approx 
\ln\min\{n_\E, n_\H(t) \, n_\R(t) \}.
\end{equation}
But we have $n_\H(t) \, n_\R(t) = n_\mathrm{H_0}  \leq n_\E,$
so 
\begin{equation}
\langle \hat S_\E(t)\rangle=  \langle \hat S_\mathrm{HR}(t)\rangle\approx 
\ln n_\mathrm{H_0} \approx S_\mathrm{Bekenstein,0}.
\end{equation}
That is, $\langle \hat S_\E(t)\rangle$ is not the total entropy of the rest of the universe; it is merely the extent to which the rest of the universe is entangled with the HR subsystem, which in turn is equal to the initial Bekenstein entropy of the black hole. While $\langle \hat S_\E(t)\rangle$ is by construction fixed and time independent, the fact of its existence is nevertheless crucial to a deeper understanding of entropy fluxes. 

\subsection{Tripartite entanglement: Mutual information}

For the more interesting tripartite system we have
\begin{equation}
I_{\H:\R} = S_\H +S_\R - S_{\H\R} = S_\H +S_\R - S_{\E}. 
\end{equation}
Now averaging over the pure states we have at all times (the argument $t$ is suppressed for clarity)
\begin{equation}
\langle \hat I_{\H:\R} \rangle = \hat S_{n_\H,n_\R n_\E} +\hat S_{n_\R,n_\H n_\E} - \hat S_{n_\E,n_\H n_\R}.
\end{equation}
But $n_\H \leq n_\R n_\E$, while $n_\R \leq n_\H n_\E$, and $n_\H n_\R\leq n_\E$.
So in dimensionless units, and using the harmonic numbers $H_n$  we have the exact result
\cite{ana:pra} 
\begin{eqnarray}
\hspace{-10pt}
\langle \hat I_{\H:\R} \rangle 
&=& \left[ H_{n_\H n_\R n_\E} - H_{n_\R n_\E} - {n_\H-1\over 2n_\R n_\E} \right] 
\nn \\
&& 
+ \left[ H_{n_\H n_\R n_\E} - H_{n_\H n_\E} - {n_\R-1\over 2n_\H n_\E} \right]
\nonumber\\
&&
- \left[ H_{n_\H n_\R n_\E} - H_{n_\E} - {n_\H n_\R-1\over 2n_\E} \right].\quad
\end{eqnarray}
Then after a little simplification
\begin{eqnarray}
\langle \hat I_{\H:\R} \rangle 
&=&  H_{n_\H n_\R n_\E} + H_{n_\E}  - H_{n_\R n_\E} - H_{n_\H n_\E}  \nn \\
&& + { (n_\H-1)(n_\R-1)(n_\H n_\R +n_\H + n_\R) \over 2n_\H n_\R n_\E}.
\nonumber
\\
&&
\end{eqnarray}
It is now relatively easy to see \cite{ana:pra} that
\begin{equation}
\langle \hat I_{\H:\R} \rangle \leq 
{ n_\H n_\R \over 2 n_\E}  = { n_{\H_0}  \over 2 n_\E}  \leq {1\over2}.
\end{equation}
So, the average mutual information between the black hole and the Hawking radiation never exceeds ${1\over2}$ nat throughout the entire evaporation process.

\subsection{Tripartite entanglement: \\
Infinite-dimensional environment}
For the bipartite HR system, the whole point is to keep the total dimensionality fixed. 
For the tripartite HRE system however, the environment is used to initially entangle the black hole with the rest of the universe, but then  ``comes along for the ride''. There is no real loss of generality in taking the limit $n_\E\to\infty$ (arbitrarily high dimensional Hilbert space). This is not making any assumptions concerning the actual thermodynamic entropy of the rest of the universe. Under these conditions we have demonstrated (at all times) \cite{ana:pra} the following limits:
\begin{eqnarray}
\lim_{n_\E\to\infty}  \langle \hat S_\H \rangle &=& \ln n_\H; \qquad \lim_{n_\E\to\infty}  \langle \hat S_\R \rangle = \ln n_\R; \nn \\
\lim_{n_\E\to\infty}  \langle \hat S_\E \rangle &=& \ln (n_\H n_\R) = \ln n_{\H_0}.
\end{eqnarray}
In this limit we therefore have the \emph{equality}
\begin{equation}
\lim_{n_\E\to\infty}  \left( \vphantom{\Big{|}} \langle S_\H \rangle +  \langle S_\R \rangle \right)
= \lim_{n_\E\to\infty}  \langle S_\E \rangle,
\end{equation}
an equality which (in this limit) reproduces the classical thermodynamic arguments, (balancing Bekenstein entropy versus Clausius entropy), that we started with.  
An immediate consequence of this result is
\begin{equation}
\lim_{n_\E\to\infty} \langle I_{\H:\R} \rangle = 0.
\end{equation}
For an infinite dimensional environment the mutual information between the subsystems H and R in a pure-state  HRE system is zero. The fact that things simplify so nicely for an infinite dimensional environment should  perhaps not be all that surprising in view of the fact that even in purely classical thermodynamics an infinite volume limit (infinite degrees of freedom) is necessary for the existence of phase transitions. In counterpoint, an infinite dimensional environment is also necessary if for some reason one wishes to drive the Shannon entropy to infinity~\cite{shannon, zipf}.

\vspace{-10pt}
\section{Discussion}

Since we know that there is no information puzzle in burning a lump of coal~\cite{burning},  or in the Hawking emission from analogue black holes~\cite{Unruh:1981,Visser:1993,Visser:1998,Visser:1998b,Visser:2001a,LRR,Barcelo:2006,Visser:2007,Visser:2010,Lake-Como,Weinfurtner:2010,Weinfurtner:2013,Unruh:amplifier,Unruh:measured,Steinhauer:2014,Steinhauer:2015a,Steinhauer:2015b,Belgiorno:2010,Schutzhold:2010,Belgiorno:2010b,Rubino:2011,Liberati:2011}, we can use this as a starting point to understand what happens in a GR black hole system.

First, we explicitly calculated the classical thermodynamic (Clausius) entropy and the Bekenstein entropy. We found that they compensate perfectly,  summing to the initial Bekenstein entropy of the hole. (As of course they must, given how Bekenstein entropy was originally defined.) Once we had the classical behaviour under control, we proceeded with a quantum entropy argument based on the von~Neumann entropy, realizing that on average we need to encode $3.9\pm2.5$ bits of information per emitted quantum to preserve unitarity. 

We have developed a tripartite system in which, assuming unitarity of the evolution of the (GR black hole) + (Hawking radiation) subsystem, we showed that, as long as it is suitably embedded in a tripartite system providing an environment to entangle with, there are no unusual physical effects;  the results completely agree with the classically expected results, to within 1 nat.

(That consideration of  the ``rest of the universe'' is necessary for making sensible statements about unitarity has also been argued, in a different context, in references~\cite{ana1,ana2}.) 
In contrast, the results previously obtained by Page correspond to the choice of a ``closed box model'' which never interacts with, (or even notices), the rest of the universe.  In that model, the consideration of a simplified and idealized bipartite system gives rise to physics that is not  well-understood; such as a zero initial Bekenstein entropy and an odd entropy/information balance that is a key part of the motivation for firewalls~\cite{firewall,apologia,Chen:2015,Mathur:2015,Nomura:2012a,Nomura:2012b,Israel:2014,Albrecht:2014}. 
On the contrary, in our system the purification process can occur continuously. Specifically, the mutual information between the black hole and the Hawking radiation, (when properly interpreted as part of a tripartite system entangled with an environment), never exceeds ${1\over2}$ nat. We also show that in the limit of infinite dimension of the environment, there is no loss of generality in our argument and, moreover, the ``sum rule'' holds exactly. This result can be related with the fact that in classical thermodynamics we need an infinite volume limit for the existence of phase transitions. Overall this leads to noncontroversial and relatively boring physics --- quite similar to burning a lump of coal~\cite{burning} --- one obtains a simple cascade of Hawking quanta~\cite{sparsity,Visser:1992}.

\vspace{-10pt}
\section*{Acknowledgements}
\noindent
AA-S is supported by the grant GACR-14-37086G of the Czech Science Foundation.
MV is supported by the Marsden Fund, administered by the Royal Society of New Zealand.

\end{document}